\documentclass[12pt]{article}

\usepackage{a4wide}
\usepackage{amsmath}
\usepackage{amsfonts}
\usepackage{amsthm}
\usepackage{dsfont}

\newcommand{\ud}{\mathrm{d}}
\newcommand{\ui}{\mathrm{i}}
\newcommand{\ue}{\mathrm{e}}

\newcommand{\R}{\mathds{R}}
\newcommand{\Z}{\mathds{Z}}
\newcommand{\N}{\mathds{N}}

\newcommand{\cS}{{\mathcal S}}

\newcommand{\cH}{{\cal H}}

\newcommand{\supp}{\operatorname{supp}}

\providecommand{\norm}[1]{\lVert#1\rVert}
\providecommand{\abs}[1]{\lvert#1\rvert}
\providecommand{\biggabs}[1]{\bigg\lvert#1\bigg\rvert}
\providecommand{\bigabs}[1]{\big\lvert#1\big\rvert}

\newcommand{\pa}{\partial}
\newcommand{\la}{\langle}
\newcommand{\ra}{\rangle}

\newcommand{\Op}{\operatorname{Op}}
\newcommand{\Tr}{\operatorname{Tr}}
\newcommand{\vol}{\operatorname{vol}}

\newtheorem{thm}{Theorem}
\newtheorem{lem}{Lemma}
\newtheorem{prop}{Proposition}
\newtheorem{cor}{Corollary}

\title{Upper bounds on the rate of quantum ergodicity}

\author{Roman Schubert\thanks{School of Mathematics, University of Bristol, e-mail: roman.schubert@bristol.ac.uk}}

\date{March 16, 2005}

\begin{document}

\maketitle

\abstract{We study the semiclassical behaviour of eigenfunctions of quantum systems 
with ergodic classical limit. By the  quantum ergodicity theorem almost all of these eigenfunctions 
become equidistributed in a weak sense. 
We give a simple derivation of
 an upper bound of order $\abs{\ln\hbar}^{-1}$ on the rate of quantum ergodicity 
if the classical system is ergodic with a certain rate. 
In addition we obtain a similar bound on transition amplitudes if the classical system is weak mixing. 
Both results generalise previous ones by Zelditch. 
We then extend the results to some classes of quantised maps on the torus and 
obtain a logarithmic rate for perturbed cat-maps and a sharp algebraic  rate for parabolic maps.}

\section{Introduction}

The quantum ergodicity theorem by 
Shnirelman, Zelditch and Colin de Verdi{\`e}re, \cite{Shn74,Zel87,Col85}, states 
that almost all eigenfunctions of a quantum mechanical Hamilton operator become equidistributed 
in the semiclassical limit if the underlying 
classical system is ergodic. 

Consider as example an Hamiltonian 
of the form 
\begin{equation}
\cH= -\hbar^2 \Delta +V
 \end{equation}
on $L^2(\R^d)$ with a smooth potential satisfying 
$\abs{\pa^{\alpha}V(x)}\leq C_{\alpha}(1+\abs{x}^2)^{m/2}$ 
for some $m\in \R$ and all 
$\alpha\in \N^d$. 
Assume that for a fixed  energy $E$ the 
classical energy-shell 
$\Sigma_E:=\{(\xi,x)\in \R^d\times \R^d\, ;\, \xi^2+V(x)=E\}$ 
is compact,  then the spectrum of $\cH$ is discrete in a neighbourhood of 
$E$, and we will denote by $N(I(E,\hbar))$ the number of eigenvalues in the interval 
$I(E,\hbar):=[E-\alpha\hbar,E+\alpha\hbar]$, $\alpha >0$.
If now the Hamiltonian flow generated by $H=\xi^2+V(x)$  ergodic on $\Sigma_E$ then the 
normalised eigenfunctions 
$\psi_n$ of $\cH$  satisfy 
\begin{equation}\label{eq:hmr}
\lim_{\hbar\to 0}\frac{1}{N(I(E,\hbar))}\sum_{E_n\in I(E,\hbar)} \abs{\la \psi_n,\Op[a]\psi_n\ra-\overline{a}_E}^2=0
\end{equation}
with $\overline{a}_E:=\frac{1}{\vol(\Sigma_E)}\int_{\Sigma_E} a\,\, \ud \mu_E$
 and where $a$ is a smooth bounded function on phase space and $\Op[a]$ its 
Weyl quantisation (defined below in \eqref{eq:weyl-quant}).  
This result is the semiclassical version of the 
quantum ergodicity theorem, which was derived in \cite{HelMarRob87}. 
It implies that almost all of the expectation values $\la \psi_n,\Op[a]\psi_n\ra$ tend to 
$\overline{a}_E$ in the limit $\hbar\to 0$, so in this sense the 
eigenfunctions  
become equidistributed on the energy-shell.

Our aim is to derive an upper bound on the rate by which the 
left hand side of \eqref{eq:hmr} approaches zero. For the eigenfunctions of the Laplacian on 
manifolds of negative curvature such a bound has 
been derived by Zelditch \cite{Zel94a}. The bound we give is of the same order, so we do not get an 
improvement on the rate, but the advantage of our method is that it is simpler and 
uses only ergodicity with a certain rate as condition on the classical flow. 
Therefore it applies to a larger class of systems. The main  input 
in the proof is the result on the semiclassical propagation of observables 
up to Ehrenfest time, \cite{BamGraPau99,BouRob02}.

We will now describe the classes of Hamiltonians and observables we 
consider, see, e.g., \cite{DimSjo99} for more details.  
We say $a(\hbar,x,\xi)\in S^m$ for $m\in \R$ if 
$a$ is smooth, satisfies 
\begin{equation}\label{eq:symb-est}
\abs{\pa^{\gamma}_{x,\xi}a(\hbar,x,\xi)}\leq C_\gamma (1+\abs{x}^2+\abs{\xi}^2)^{m/2} 
\end{equation}
for all $\gamma\in \N^{2d}$ and $\hbar\in (0,1/2]$, and 
has an asymptotic expansion $a(\hbar,x,\xi)\sim \sum_{n\in \N} \hbar^n a_n(x,\xi)$, i.e., $(a-\sum_{n=0}^{N-1} \hbar^na_n)\hbar^{-N}$ satisfies \eqref{eq:symb-est} for all $N\in\N$. 
Now let $M$ be a smooth manifold, the set of operators $\Psi^m(M)$ is  given by local Weyl quantisation 
of these classes, if $a\in S^m$ in some local chart, then $\Op[a]$ is defined as 
\begin{equation}\label{eq:weyl-quant}
\Op[a]\psi=\frac{1}{(2\pi\hbar)^d} \iint \ue^{\frac{\ui}{\hbar} \la x-y,\xi\ra}a\big(\hbar,\frac{x+y}{2},\xi\big) \psi(y)\,\, \ud y\ud\xi\,\, .
\end{equation}
A general operator $A\in \Psi^m(M)$ is then an operator who is locally of the form 
\eqref{eq:weyl-quant} with some $a\in S^m$. The function $a$ is called the local 
symbol of the operator $A$ and the leading term in the asymptotic expansion of $a$ is called 
the principal symbol 
\begin{equation}
\sigma(A):=a_0\,\, ,
\end{equation}
the principal symbol can be glued together to  a function on $T^*M$, but the full symbol  not. 
The operators in $\Psi^0(M)$ are bounded on $L^2(M)$ (uniformly in $\hbar$) and will form 
our basic class of observables.

We will assume that the Hamiltonian $\cH$ is a selfadjoint operator in $\cH\in \Psi^m(M)$, for some 
$m >0$, and denote by $\Phi^t$ the Hamiltonian flow on $T^*M$ generated 
by the principal symbol $H_0=\sigma(\cH)$ of $\cH$. 
Let $\Sigma_E:=\{ (x,\xi)\in T^*M\, ;\, H_0(x,\xi)=E\}\subset T^*M$ 
denote the energy surface and $\ud\mu_E$ the Liouville measure on 
$\Sigma_E$.  If $E$ is 
a regular value of $H_0$ and $\Sigma_{E}$ is compact, 
 then the spectrum of $\cH$ is discrete in a 
neighbourhood of $E$. If furthermore the set of periodic orbits of $\Phi^t$ on 
$\Sigma_E$ has measure zero, then the number of eigenvalues close to $E$ 
satisfies   the Weyl estimate 
\begin{equation}\label{eq:weyl}
N(I(E,\hbar))=\frac{2\alpha}{(2\pi)^d\hbar^{d-1}} \vol(\Sigma_E)(1+o(1))\,\, ,
\end{equation}
where $\vol(\Sigma_E):=\int_{\Sigma_E}\,\ud\mu_E$ and 
$\ud \mu_E$ denotes the Liouville measure on $\Sigma_E$, see 
\cite{PetRob85,Ivr98,DimSjo99}.

The autocorrelation function at energy $E$ of a function $a$ on $T^*M$ 
is defined as 
\begin{equation}
C_E[a](t):=\frac{1}{\vol(\Sigma_E)}\int_{\Sigma_E}a\circ \Phi^t a\, \ud \mu_E-
\big(\overline{a}_E\big)^2\,\, ,
\end{equation} 
where 
\begin{equation}\label{eq:def-mean}
\overline{a}_E:=\frac{1}{\vol(\Sigma_E)}\int_{\Sigma_E}a\, \ud\mu_E\,\, .
\end{equation}
The flow $\Phi^t$ is  ergodic on $\Sigma_E$ if for every  $a\in L^1(\Sigma_E,\ud\mu_E)$ one has 
\begin{equation}
\lim_{T\to\infty}\frac{1}{T}\int_0^T C_E[a](t)\, \ud t=0\,\, , 
\end{equation}
see \cite{Wal82}. 
We will say that $\Phi^t$ is ergodic with 
rate $\gamma>0$ on $\Sigma_E$ if for every 
$a\in C^{\infty}(\Sigma_E)$ and $f\in \cS(\R)$  there is a constant 
$C$ such that 
\begin{equation}\label{eq:mixing-rate}
\frac{1}{T}\int f\bigg(\frac{t}{T}\bigg) C_E[a](t)\,\, \ud t\leq C (1+\abs{T})^{-\gamma}\,\, .
\end{equation} 
The rate of ergodicity can be related to the more  common rate of mixing, 
the system is called mixing if $\lim_{t\to\infty} C_E[a](t)=0$,  and if 
$\abs{C_E[a](t)}\leq C(1+\abs{t})^{-\tilde{\gamma}}$, then $\tilde{\gamma}$ is called the
rate of mixing.  We see from \eqref{eq:mixing-rate} that for $0<\tilde{\gamma}<1$ we have at least a rate 
of ergodicity 
$\gamma=\tilde{\gamma}$, whereas for $\tilde{\gamma}>1$ we  have at least $\gamma=1$. So a rate 
of mixing implies a 
rate of ergodicity, but the contrary is not true, there are systems which are not mixing but which can have 
a large rate of ergodicity due to an oscillatory behaviour of $C_E[a](t)$. Examples are easily found 
among maps, for instance the Kronecker map, and we will discuss some cases in the last section about 
quantised maps.

Our main result is now 

\begin{thm}\label{thm:rate}
Let $\cH\in \Psi^m(M)$, for some $m>0$, be selfadjoint with principal symbol 
$H_0$. Assume that $E$ is a regular value of $H_0$, that 
$\Sigma_E$ is compact and denote by $E_n$, $\psi_n$ the eigenfunctions and 
eigenvalues of $\cH$ in the interval $I(E,\hbar)=[E-\alpha \hbar,E+\alpha\hbar]$, $\alpha>0$. If the Hamiltonian flow 
$\Phi^t$ generated by $H_0$ is  
 ergodic with rate $\gamma>0$ on $\Sigma_E$, then for every 
$A\in \Psi^0(M)$ there exists a $C>0$ such that 
\begin{equation}\label{eq:bound}
\frac{1}{N(I(E,\hbar))}\sum_{E_n\in  I(E,\hbar)}
\abs{\la \psi_n, A\psi_n\ra -\overline{\sigma(A)}_E}^2\leq 
C \begin{cases}  \abs{\ln\hbar}^{-\gamma} & \text{if} \,\, 0<\gamma\leq 1 \\
\abs{\ln\hbar}^{-1} & \text{if} \,\, \gamma \geq1 \end{cases}\,\, ,
\end{equation}
where $\overline{\sigma(A)}_E$ is defined in \eqref{eq:def-mean}.
\end{thm}

This result is an extension of the previous result by Zelditch, \cite{Zel94a}, 
who obtained the same logarithmic bound for $\gamma>1$ for eigenfunctions of 
the Laplacian on compact manifolds 
of negative 
curvature (in order to connect the two setups one has to rescale the Laplacian with  $\hbar$). 
The improvement lies in the weakening of 
 the assumptions to a rate of ergodicity and in 
a simpler proof, this is possible 
because we can use the recent results on propagation 
of observables up to Ehrenfest time \cite{BamGraPau99,BouRob02}. 
A similar result has been stated recently by Robert in the review \cite{Rob04}. 

Further systems where Theorem \ref{thm:rate} 
applies are Schr{\"o}dinger operators $\cH=-\hbar^2\Delta +V$ on the $2$-torus with the 
smooth potentials $V$,  constructed by Donnay and Liverani 
\cite{DonLiv91}, for which the flow is ergodic and mixing \cite{BalTot03}.  
These examples have been recently generalised to higher dimensions, \cite{BalTot05}.

For strongly chaotic systems the bound \eqref{eq:bound} 
is far from the conjectured 
optimal one. For eigenfunctions of the Laplace Beltrami operator  on compact 
surfaces of negative curvature, where the corresponding classical system is the 
the geodesic flow, which is Anosov,   Rudnick and Sarnak 
\cite{RudSar94,Sar03} 
have conjectured that  
\begin{equation}
\biggabs{\la \psi_n, \rho\psi_n\ra -\int\rho\, \ud \nu_g}\leq C_{\varepsilon}E_n^{-1/4+\varepsilon}
\end{equation}
holds for all $\varepsilon>0$. Here $\rho$ is a sufficiently nice function on the surface  
and $\ud \nu_g$ is the Riemannian volume element. 
Translated in our context that would imply 
a bound $h^{1-\varepsilon}$ in \eqref{eq:bound}. A very precise prediction 
for the behaviour of the sum on the left hand side of \eqref{eq:bound} 
has been derived in \cite{EckFisKea95}, for a compact uniformly hyperbolic system
with time reversal invariance and no other symmetry it reads
\begin{equation}\label{eq:second-moment}
\begin{split}
\frac{1}{N(I(E,\hbar))}\sum_{E_n\in  I(E,\hbar)}&
\abs{\la \psi_n, A\psi_n\ra -\overline{\sigma(A)}_E}^2\\
&=2\frac{(2\pi\hbar)^{d-1}}{\vol \Sigma_E}\int_{-\infty}^{\infty}C_E[\sigma(A)](t)\,\, \ud t+o(\hbar^{d-1})\,\, .
\end{split}
\end{equation}

These predictions have been numerically tested in \cite{EckFisKea95, AurTag98,BaeSchSti98}, and confirmed for uniformly hyperbolic 
systems like manifolds of negative curvature. For non-uniformly 
hyperbolic systems like Euclidean billiards the findings are less clear and  the rate is sometimes slower, at least in the tested energy range. So understanding the rate 
of quantum ergodicity remains a major open problem. 
Very recently Luo and Sarnak, see \cite{Sar03}, established a result 
of the form \eqref{eq:second-moment} for the discrete spectrum of the 
Laplacian on the modular surface. But due to the  arithmetic 
nature of the system the right hand side of \eqref{eq:second-moment} differs and an  additional factor related to $L$-functions 
appears.

The reason for the rather large gap between the estimate \eqref{eq:bound} and the conjectured one is 
our poor understanding of the quantum time evolution for large times when the underlying classical 
system is hyperbolic. In our present techniques the hyperbolicity leads to exponentially growing 
remainder terms and this reduces us to time scales which are logarithmic in $\hbar$. 
But for systems which are ergodic but not hyperbolic we can hope to get much stronger results. 
Examples for such systems can be constructed as maps on the torus, and we therefore have added 
a section on quantised maps. In this section we will first prove  an analogue of Theorem 
\ref{thm:rate} for perturbed cat maps using techniques from \cite{BouDeB04}, and then we 
study the quantised parabolic map introduced in \cite{MarRud00} and show that we get 
an algebraic decay of \eqref{eq:bound}, with an optimal rate.

The method we use to prove Theorem \ref{thm:rate} can be used as well 
to get a bound on the off-diagonal matrix elements. We say that the flow $\Phi^t$ is weak mixing with rate $\gamma >0$ 
on $\Sigma_E$ if for all smooth $a$ on $\Sigma_E$ and $f\in \cS(\R)$ there is a constant $C$ such that for all 
$\varepsilon \in\R$ 
\begin{equation}\label{eq:rate-weak-mixing}
\frac{1}{T}\int f\bigg(\frac{t}{T}\bigg)C_E[a](t)\ue^{\ui \varepsilon t}\,\, \ud t \leq C(1+\abs{T})^{-\gamma}\,\, .
\end{equation}
That the above quantity tends to $0$ for $T\to\infty$ is equivalent to weak mixing, so the above condition quantifies 
the rate of weak mixing. As for the rate of ergodicity, a rate of mixing implies a similar rate of weak mixing.

\begin{thm}\label{thm:off-diag}
Under the same conditions as in Theorem \ref{thm:rate} we have for 
$\gamma>0$ 
\begin{equation}
\frac{1}{N(I(E,\hbar))}
\sideset{}{'}\sum_{\begin{subarray}{l} n,m \, ;\, E_n\in  I(E,\hbar)\\
\abs{E_n-E_m}\leq \hbar/\abs{\ln\hbar}\end{subarray}}
\abs{\la \psi_n, A\psi_m\ra}^2\leq  C \begin{cases}  \abs{\ln\hbar}^{-\gamma} & \text{if} \,\, 0<\gamma\leq 1 \\
\abs{\ln\hbar}^{-1} & \text{if} \,\, \gamma \geq1 \end{cases}\,\, ,
\end{equation}
and if the flow is weak mixing with a rate $\gamma>0$, then for any $\varepsilon\in\R$
\begin{equation}\label{eq:off-diag-weak-mix}
\frac{1}{N(I(E,\hbar))}
\sideset{}{'}\sum_{\begin{subarray}{l} n,m \, ;\, E_n\in  I(E,\hbar)\\
\abs{E_n-E_m-\hbar \varepsilon}\leq \hbar/\abs{\ln\hbar}\end{subarray}}
\abs{\la \psi_n, A\psi_m\ra}^2\leq  C \begin{cases}  \abs{\ln\hbar}^{-\gamma} & \text{if} \,\, 0<\gamma\leq 1 \\
\abs{\ln\hbar}^{-1} & \text{if} \,\, \gamma \geq1 \end{cases}  \,\, ,
\end{equation}
where the prime at the sum indicates that we sum over $E_m$, $E_n$ with 
$E_m\neq E_n$.
\end{thm}

The behaviour of off-diagonal matrix elements has been studied by Zelditch 
\cite{Zel90,Zel96}  who showed that ergodicity and weak mixing implies that the above sums 
tend to zero for $\hbar\to 0$. Further results have been derived in 
\cite{Tat99}. As we will see in Section \ref{sec:maps}  weak mixing is a necessary condition for 
\eqref{eq:off-diag-weak-mix} to hold.

The plan of the paper is as follows. The next two sections are devoted to the proof 
of Theorems  \ref{thm:rate} and \ref{thm:off-diag}. In section \ref{sec:prel} we collect 
some preliminaries, and in section \ref{sec:proofs} we do the proofs. 
In the final section \ref{sec:maps} we then discuss some quantised maps.

{\bf Acknowledgements:} This work has been  supported by the European 
Commission under the Research Training Network 
(Mathematical Aspects of Quantum Chaos) 
$\rm{n}^{o}$ HPRN-CT-2000-00103 of the IHP Programme.


\section{Preliminaries}\label{sec:prel}

The proofs of Theorems \ref{thm:rate} and \ref{thm:off-diag} rest on two ingredients, a microlocal version of 
Weyl's law and 
a version of Egorov's theorem which is valid up to Ehrenfest time. 
In this section we will recall these results and present them in the 
form we need.

The estimates collected in this section will be finally applied to  compute 
\begin{equation}
\Tr \rho\big((E-\cH)/\hbar\big)B U^*(t)AU(t) 
\end{equation}
for $A,B\in \Psi^0(M)$. This quantity can be localised by splitting $B=\sum_j \Op[b_j]$ with 
$b_j$ supported (modulo $\hbar^{\infty}$) in local charts. Therefore it is sufficient for us to work 
in $M=\R^d$, and this will facilitate  some of the remainder estimates.

For a function $a\in C^{\infty}(\R^m)$ we will use the notation 
\begin{equation}
\abs{a}_k:=\sum_{\abs{\alpha}\leq k}\sup_{x\in\R^m} \abs{\pa^{\alpha}a(x)}
\end{equation}
for $k\in\N$.

\begin{prop}\label{prop:local-Weyl}
Assume that $\cH\in \Psi^m$ is selfadjoint and has principal symbol 
$H_0$. Assume furthermore that $E$ is a regular value of $H_0$ and that  
$\Sigma_E$ is compact. Let $\rho$ be a smooth function on $\R$ such 
that the Fourier transform $\hat{\rho}$ 
has compact support in a small neighbourhood of $0$ which contains no period 
of a periodic orbit of $\Phi^t$ on $\Sigma_E$.  Then there is a constant  
$C>0$ such that for every $\Op[b]\in \Psi^0$ we have 
\begin{equation}\label{eq:local-Weyl}
\biggabs{\sum_{E_n} \rho\bigg(\frac{E-E_n}{\hbar}\bigg) \la \psi_n, \Op[b]\psi_n\ra- 
\frac{\hat{\rho}(0)}{(2\pi)^d\hbar^{d-1}}\, \overline{\sigma(b)}_E}
\leq C\hbar^{2-d}  \abs{\rho}_5\abs{b}_{2d+8}\,\, .
\end{equation}
\end{prop}

The proposition is a standard result and well known in literature, 
except that the way that the error term depends on 
$b$ is usually not made explicit. Since the main tool in deriving 
the formula  \eqref{eq:local-Weyl} is the method of stationary phase, 
or variants thereof, it comes as no surprise that the error term 
can be estimated by a finite number of derivatives of $b$. 
An analogous result for high-energy asymptotics on compact manifolds 
was derived in \cite{Zel94a}. 
For  convenience  we will sketch the proof 
of Proposition \ref{prop:local-Weyl}, for details we frequently  refer to \cite{DimSjo99}. . 

\begin{proof}
We first observe that without loss of generality we can assume that
$b$ is supported in a compact neighbourhood of the energy-shell 
$\Sigma_E$.  Let $f(E)$ be a smooth function with compact support 
such that $f(H(x,\xi))$ has compact support and 
$f(H(x,\xi))\equiv 1$ on a neighbourhood of $\Sigma_E$. 
By the functional calculus 
one has then $f(\cH)\in\Psi(1)$, see \cite{DimSjo99}. 
Let $U(t)=\ue^{-\frac{\ui}{\hbar} t\cH}$ be the time evolution operator, i.e., 
the solution to $\ui\hbar \pa_t U(t)=\cH U(t)$ with initial condition $U(0)=I$. 
One then  constructs an approximation to the operator $U_f(t)=U(t)f(\cH)$ by solving 
the initial value problem 
\begin{equation}
\big(\ui\hbar \pa_t-\cH\big) U_f(t)=0\,\, ,\quad  U_f(0)=f(\cH)
\end{equation}
approximately for small $t$, i.e., for every $N\in \N$ one can find an $V^{(N)}(t)$ such that 
\begin{equation}\label{eq:aprox-prop}
\big(\ui\hbar \pa_t-\cH\big) V^{(N)}(t)=\hbar^{N+1} R_N(t)\,\, ,\quad  V^{(N)}(0)=f(\cH)\,\, ,
\end{equation}
with $\norm{R_N(t)}\leq C$ for $t\in [-T_0,T_0]$ where $T_0$ is smaller then the period of 
the shortest periodic orbit 
on $\Sigma_E$. Then Duhamel's principle gives 
\begin{equation}
 U_f(t)=V^{(N)}(t)+\ui\hbar^N \int_0^t U_f(t-t')R_N(t')\,\, \ud t'
\end{equation}
and therefore 
\begin{equation}
\begin{split}
\abs{\Tr U_f(t)\Op[b]-\Tr V^{(N)}(t)\Op[b]}&\leq \hbar^N \abs{t} \sup_{t'\in[0,t]} \abs{\Tr U_f(t-t')R_N(t') \Op[b]}\\
&\leq \hbar^N  C_N \Tr \abs{ \Op[b]}
\end{split}
\end{equation}
since $ \abs{t} \sup_{t'\in[0,t]} \norm{U_f(t-t')R_N(t')}\leq C$ for $t\in[-T_0,T_0]$ and 
we have used the general relation 
$\abs{\Tr AB}\leq \norm{A}\Tr \abs{B}$ if $A$ is bounded and $B$ of trace class.  
Since $b$ is of compact support $\Op[b]$ is of trace class and its 
trace norm can be estimated as 
\begin{equation}
\Tr \abs{ \Op[b]}\leq C\frac{1}{(2\pi\hbar)^d}\, 
\abs{b}_{2d+1}\,\, ,
\end{equation}
see \cite[Chapter 9]{DimSjo99}. 
The kernel of $V^{(N)}(t)$ satisfying  \eqref{eq:aprox-prop} is given by 
\begin{equation}\label{eq:aprop}
V^{(N)}(t,x,y)=\frac{1}{(2\pi\hbar)^d}\int \ue^{\frac{\ui}{\hbar}[\varphi(t,x,\xi)-y\xi]} a^{(N)}(t,x,\xi)\,\, \ud\xi 
\end{equation}
where $\varphi(t,x,\xi)$ is a solution to the Hamilton Jacobi equation 
\begin{equation}\label{eq:ham-jac}
\pa_t\varphi(t,x,\xi) + H(x,\varphi_x'(t,x,\xi))=0
\end{equation}
with initial condition $\varphi(0,x,\xi)=x\xi$, and 
$a^{(N)}(t,x,\xi)\in C^{\infty}([-T_0,T_0],S^1)$ is the solution 
of a corresponding transport equation with initial condition 
$a^{(N)}(0,x,\xi)=f(H(x,\xi))+O(\hbar)$ 
given by the symbol of $f(\cH)$. See \cite[Chapter 10]{DimSjo99} for 
the proof and more details. 
If $\tilde{b}=\ue^{\ui \hbar \pa_x\pa_{\xi}}b$ denotes the 
left 
symbol of $\Op[b]$ (the case $t=0$ in \cite[Equation (7.5)]{DimSjo99}) 
then we get from \eqref{eq:aprop} 
\begin{equation}
\begin{split}
\int \ue^{\frac{\ui}{\hbar} Et}& \Tr\big[V^{(N)}(t)\Op[b]\big]\hat{\rho}(t)\,\, \ud t \\
&\quad = 
\frac{1}{(2\pi\hbar)^d}\iiint \ue^{\frac{\ui}{\hbar}[\varphi(t,x,\xi)-x\xi+Et]} \hat{\rho}(t)a^{(N)}(t,x,\xi)\tilde{b}(x,\xi)\, \,\ud x\ud\xi\ud t\,\, .
\end{split}
\end{equation}
The main contributions to this integral come from the points where the phase is stationary, 
the stationary phase 
condition reads 
\begin{equation}
\pa_t\varphi(t,x,\xi)+E=0\,\, ,\quad 
\pa_x\varphi(t,x,\xi)-\xi=0\, \quad \text{and}\quad 
\pa_\xi\varphi(t,x,\xi)-x=0\,\, .
\end{equation}
In view of \eqref{eq:ham-jac} the first equation means that $H(x,\xi)=E$ 
and the second and third 
mean that $\Phi^t(x,\xi)=(x,\xi)$, i.e., $(x,\xi)$ lie on a periodic orbit with 
period $t$. Since by assumption 
the support of $\hat{\rho}$ does not contain any period of a periodic orbit, 
the only stationary points left 
are at $t=0$, and consist of the whole energy shell $\Sigma_E$. 
Because $E$ is assumed to be 
a non-degenerate energy level we can choose new coordinates $(E',z)$ in a 
neighbourhood of $\Sigma_E$ such that $H(E',z)=E'$, when we use 
furthermore that $\varphi(t,x,\xi)=x\xi-tH(x,\xi)+r(t,x,\xi)$ with 
$r(t,x,\xi)=O(t^2)$ which follows from \eqref{eq:ham-jac}, 
then the above integral becomes 
\begin{equation}
\frac{1}{(2\pi\hbar)^d}\iiint \ue^{\frac{\ui}{\hbar}[(E-E')t+r(t,E',z)]} \hat{\rho}(t)a^{(N)}(t,E',z)\tilde{b}(E',z)J(E',z)\, \,\ud E'\ud t\ud z\,\, ,
\end{equation}
where $J(E',z)$ denotes the Jacobian of the change of coordinates. We can now apply the 
stationary phase theorem with remainder estimate to the $t,E'$ integrals and get 
\begin{equation}
\begin{split}
\frac{1}{2\pi\hbar}\iint \ue^{\frac{\ui}{\hbar}[(E-E')t+r(t,E',z)]} \hat{\rho}(t) & a^{(N)}(t,E',z)\tilde{b}(E',z)J(E',z)\, \,\ud E'\ud t\\
& =\hat{\rho}(0)a^{(N)}(0,E,z)\tilde{b}(E,z)J(E,z)+O(\hbar \abs{\rho}_5\abs{\tilde{b}}_5)\,\, ,
\end{split}
\end{equation}
where the implied constant does only depend on $a$ and $\varphi$. 
With the initial condition 
$a^{(N)}(0,E,z)=1+(\hbar^\infty)$ and $\abs{\pa^{\alpha}b-\pa^{\alpha}\tilde{b}}\leq C\abs{b}_{\abs{\alpha}+2d+3}$ 
we then finally obtain
\begin{equation}
\biggabs{\int \ue^{\frac{\ui}{\hbar} Et}\Tr(V^{(N)}(t)\Op[b])\hat{\rho}(t)\,\, \ud t-
\frac{\hat{\rho}(0)}{(2\pi\hbar)^{d-1}}\int_{\Sigma_E}\sigma(b)\,\, \ud\mu_E}\leq C\hbar^{d-2} \abs{\rho}_5\abs{b}_{2d+8} \,\, .
\end{equation}

On the other hand side, by the spectral resolution of $U(t)$ we have 
\begin{equation}
\int \ue^{\frac{\ui}{\hbar} Et}\Tr(U_f(t)\Op[b])\hat{\rho}(t)\,\, \ud t 
=2\pi \sum_{E_n}\rho\bigg(\frac{E-E_n}{\hbar}\bigg) \la \psi_n, \Op[b]\psi_n\ra 
\end{equation}
and so finally we get 
\begin{equation}
\begin{split}
\sum_{E_n}\rho\bigg(\frac{E-E_n}{\hbar}\bigg)& \la \psi_n, \Op[b]\psi_n\ra\\ 
&=\frac{\hat{\rho}(0)}{(2\pi)^d\hbar^{d-1}}\overline{\sigma(b)}_E
+O(\hbar^{d-2} \abs{\rho}_5\abs{b}_{2d+8} )+O(\hbar^{d-N}\abs{\rho}_0\abs{b}_{2d+1})
\end{split}
\end{equation}
where the implied constants do only depend on $a$, $\varphi$ and $f$. 
\end{proof}

We want to use this Proposition with $\Op[b]=\Op[a]U^*(t)\Op[a]U(t)$ 
where $\Op[a]\in \Psi^0$. 
In order to do so we will use the Theorem of Egorov with remainder estimate 
from \cite{BamGraPau99} and \cite[Proposition 2.7]{BouRob02}. 

\begin{thm}[\cite{BouRob02}]\label{thm:egorov}
Assume that $\cH\in\Psi^m$ is selfadjoint and let $U(t):=\ue^{-\frac{\ui}{\hbar}t\cH}$. 
Then for any compact $\Omega\subset \R^d\times \R^d$ there exists a constant  
$\Gamma_1>0$  such that  for every $\Op[a]\in\Psi^0$ with $\supp a\subset \Omega$ 
there is a $C>0$ with   
\begin{equation}
\norm{U^*(t)\Op[a]U(t)-\Op[a\circ\Phi^t]}\leq C \hbar \ue^{\Gamma_1 \abs{t}}
\end{equation}
\end{thm}

From this we get 

\begin{cor}\label{cor:egorov}
Under the assumption in  Theorem \ref{thm:egorov} there exists a constant  
$\Gamma>0$   such that for every $\Op[a]\in\Psi^0$  with support in $\Omega$ 
there is a $C>0$ with   
\begin{equation}
\norm{\Op[a]^*U^*(t)\Op[a]U(t)-\Op[a^* a\circ\Phi^t]}\leq C \hbar \ue^{\Gamma \abs{t}}
\end{equation}
\end{cor}

\begin{proof}
Using the triangle inequality and Egorov's theorem we get  
\begin{equation}
\begin{split}
&\norm{\Op[a]^*U^*(t)\Op[a]U(t)-\Op[a^* a\circ\Phi^t]}\\
&\qquad\qquad\leq \norm{\Op[a]^*U^*(t)\Op[a]U(t)-\Op[a]^*\Op[ a\circ\Phi^t]}\\
&\qquad\qquad\qquad+\norm{\Op[a]^*\Op[ a\circ\Phi^t]-\Op[a^* a\circ\Phi^t]}\\
&\qquad\qquad\leq C\hbar \norm{\Op[a]}  \ue^{\Gamma_1 \abs{t}}
+\norm{\Op[a]^*\Op[ a\circ\Phi^t]-\Op[a^* a\circ\Phi^t]}
\end{split}
\end{equation}
and since $\Op[a]$ is bounded we only have to estimate the 
second term. By the product formula for pseudo-differential operators 
and the Calderon Vallaincourt Theorem there exists a $k\in \N$ 
such that 
\begin{equation}
\norm{\Op[a]\Op[b]-\Op[ab]}\leq C\hbar \abs{a}_k\abs{b}_k
\end{equation}
where $C$ does not depend on $a$ and $b$. We use this estimate 
with $b=a\circ\Phi^t$ and that for some $\Gamma_k>0$ 
\begin{equation}
\abs{a\circ\Phi^t}_k\leq C\ue^{\Gamma_k \abs{t}}\,\, ,
\end{equation}
see \cite[Lemma 2.4]{BouRob02}. This proves the Corollary with 
$\Gamma=\max\{\Gamma_1,\Gamma_k\}$. 
\end{proof}

Using Corollary \ref{cor:egorov} together with Proposition 
\ref{prop:local-Weyl} we obtain

\begin{cor}\label{cor:variance}
There exists $C>0$, $\Gamma>0$ and $k\in \N$ such that for every selfadjoint $\Op[a]\in \Psi^0$ 
\begin{equation}
\begin{split}
\sum_{E_n, E_m} \rho\bigg(\frac{E-E_n}{\hbar}\bigg) &
\ue^{\frac{\ui}{\hbar} t (E_n-E_m)}
\bigabs{\la \psi_n, \Op[a]\psi_m\ra-\overline{\sigma(a)}_E}^2\\
&= 
\frac{\hat{\rho}(0)}{(2\pi)^d\hbar^{d-1}}\, C_E[\sigma(a)](t)
+O(\hbar^{2-d}  \abs{\rho}_5\abs{a}_{k}\ue^{\Gamma \abs{t}})\,\, .
\end{split}
\end{equation}
\end{cor}

This kind of relationship between transition amplitudes and the autocorrelation function 
is well known, 
the only new piece is that we have an explicit estimate on the time dependence of 
the remainder term. 
In fact if we multiply with a function $f(t)$  of compact support and 
integrate over $t$ we obtain 
\begin{equation}\label{eq:old-variance}
\begin{split}
\sum_{E_n, E_m} \rho\bigg(\frac{E-E_n}{\hbar}\bigg) &\hat{f}\bigg(\frac{E_m-E_n}{\hbar}\bigg)
\bigabs{\la \psi_n, \Op[a]\psi_m\ra-\overline{\sigma(a)}_E}^2\\
&=\frac{\hat{\rho}(0)}{(2\pi)^d\hbar^{d-1}}\, \int C_E[\sigma(a)](t)\, f(t)\,\, \ud t 
+O(\hbar^{2-d})\,\, ,
\end{split}
\end{equation}
which was derived in \cite{FeiPer86,Wil87} and proved in \cite{ComRob94}.


\section{Proofs of Theorems \ref{thm:rate} and \ref{thm:off-diag}}\label{sec:proofs}
The proof of Theorem \ref{thm:rate}
will rely on the fact that by Corollary \ref{cor:variance} we can let the support of 
$f$ in \eqref{eq:old-variance} become larger with $\hbar$. 

\begin{proof}[Proof of Theorem \ref{thm:rate}] 
We will assume in the following that $\overline{a}_E=0$, this can always 
be achieved by subtracting  $\overline{a}_E$ from $a$. Choose $\rho$ such that 
$\rho\geq 0$, $\rho(\frac{E-E'}{\hbar})\geq 1$ for $E'\in I(E,\hbar)$. 
Choose furthermore $f$ such that 
$\hat{f} \in C^\infty([-1,1])$ and  $f\geq 0$ and $f(0)=1$ 
and set $f_{T}(\tau):=f(T \tau)$ so that 
$\widehat{ f_{T}}(t)
=\hat{f}(t/T)/T$. Then we have 
\begin{equation}\label{eq:diag-nondiag}
\sum_{E_n\in I(E,\hbar)}\abs{\la \psi_n, \Op[a]\psi_n\ra}^2
\leq \sum_{E_n, E_m} \rho\bigg(\frac{E-E_n}{\hbar}\bigg) 
f_{T}\bigg(\frac{E_m-E_n}{\hbar}\bigg)
\abs{\la \psi_n, \Op[a]\psi_m\ra}^2\,\, .
\end{equation}
and with Corollary \ref{cor:variance} we get 
\begin{equation}\label{eq:rho-f}
\begin{split}
\sum_{E_n, E_m} &\rho\bigg(\frac{E-E_n}{\hbar}\bigg) 
f_{T}\bigg(\frac{E_m-E_n}{\hbar}\bigg)
\abs{\la \psi_n, \Op[a]\psi_m\ra}^2\\
&=\frac{\hat{\rho}(0)}{(2\pi)^d\hbar^{d-1}}\, \int C_E[\sigma(a)](t)\widehat{f_T}(t)\,\, \ud t
+O\bigg(\hbar^{2-d}  \abs{\rho}_5\abs{a}_{k}\int \ue^{\Gamma \abs{t}}\widehat{f_T}(t)\,\, \ud t\bigg)\,\, .
\end{split}
\end{equation}
Now we have
\begin{equation}\label{eq:rho-f-rem}
\biggabs{\int \ue^{\Gamma \abs{t}}\widehat{f_T}(t)\,\, \ud t}
\leq \abs{\hat{f}}_0 \frac{1}{\Gamma T} 
\ue^{\Gamma T} 
\end{equation}
and with \eqref{eq:mixing-rate} we obtain
\begin{equation}
\biggabs{\int C_E[\sigma(a)](t)\widehat{f_T}(t)\,\, \ud t}\leq 
\begin{cases} 
C\frac{1}{T} &\text{for}\,\, \gamma \geq1 \\ 
C\frac{1}{T^{\gamma}} &\text{for}\,\, 0<\gamma \leq1 
\end{cases} \,\, ,
\end{equation}
for large $T$, since $\overline{a}_E=0$ by assumption. If we choose 
\begin{equation}\label{eq:choice-T}
T=\frac{1}{\Gamma}\, \abs{\ln(\hbar)} 
\end{equation}
then $\hbar\ue^{\Gamma T}=1$, and therefore 
we get 
\begin{equation}\label{eq:double-sum}
\sum_{E_n, E_m} \rho\bigg(\frac{E-E_n}{\hbar}\bigg) 
f_T\bigg(\frac{E_m-E_n}{\hbar}\bigg)
\abs{\la \psi_n, \Op[a]\psi_m\ra}^2 \leq 
C \hbar^{d-1}\begin{cases}  \abs{\ln\hbar}^{-\gamma} & \text{if} \,\, 0<\gamma\leq 1 \\
\abs{\ln\hbar}^{-1} & \text{if} \,\, \gamma \geq1 \end{cases}  \, \, .
\end{equation}
Combining this inequality with the estimate \eqref{eq:diag-nondiag} and 
the asymptotic for the number of eigenvalues in 
$I(E,\hbar)$, \eqref{eq:weyl}, finally gives 
\begin{equation}\label{eq:final}
\frac{1}{N(I(E,\hbar))}\sum_{E_n\in I(E,\hbar)}\abs{\la \psi_n, \Op[a]\psi_n\ra}^2
\leq C\begin{cases}  \abs{\ln\hbar}^{-\gamma} & \text{if} \,\, 0<\gamma\leq 1 \\
\abs{\ln\hbar}^{-1} & \text{if} \,\, \gamma \geq1 \end{cases}  
\end{equation}
and the proof is complete.
\end{proof}

Theorem \ref{thm:off-diag} is proved along the same lines. 

\begin{proof}[Proof of Theorem \ref{thm:off-diag}]
The proof is based on relation \eqref{eq:double-sum}, notice that the 
only assumption on $\rho$ and $f$ which entered the 
derivation are that $\hat{f}$ has compact support and 
$\hat{\rho}$ is supported in $(-T_0,T_0)$. We choose now 
$\rho$ as before and $f$  such that 
\begin{equation}
f\geq \chi_{[-\Gamma,\Gamma]}
\end{equation} 
where $\chi_{[-\Gamma,\Gamma]}$ is the characteristic function of the interval 
$[-\Gamma,\Gamma]$. Then we get using \eqref{eq:double-sum} 
\begin{equation}
\frac{1}{N(I(E,\hbar))}\sum_{\begin{subarray}{l}n,m\, :\, E_n\in I(E,\hbar)\\ \abs{E_n-E_m}\leq \hbar/\abs{\ln\hbar}\end{subarray}} \abs{\la\psi_n\Op[a]\psi_m\ra}^2\leq  C\begin{cases}  \abs{\ln\hbar}^{-\gamma} & \text{if} \,\, 0<\gamma\leq 1 \\
\abs{\ln\hbar}^{-1} & \text{if} \,\, \gamma \geq1 \end{cases} 
\end{equation}
if $\overline{a}_E=0$. Together with \eqref{eq:final} this gives 
\begin{equation}
\frac{1}{N(I(E,\hbar))}\sideset{}{'}\sum_{\begin{subarray}{l}n,m\, :\, E_n\in I(E,\hbar)\\\abs{E_n-E_m}\leq \hbar/\abs{\ln\hbar}\end{subarray}}\abs{\la\psi_n\Op[a]\psi_m\ra}^2\leq  C\begin{cases}  \abs{\ln\hbar}^{-\gamma} & \text{if} \,\, 0<\gamma\leq 1 \\
\abs{\ln\hbar}^{-1} & \text{if} \,\, \gamma \geq1 \end{cases} 
\end{equation}
and since  $\la\psi_m, \overline{a}_E\psi_n\ra=0$ if 
$E_m\neq E_n$, this estimate is true for all $\Op[a]\in\Psi^0$. 

With the same choices of $\rho$ and $f$ and by shifting $f_T$, 
\begin{equation}
f^{(\varepsilon)}_T(\tau):=f_T(\tau-\varepsilon)\,\, ,
\end{equation} 
we get from \eqref{eq:rho-f} and \eqref{eq:rho-f-rem}    
\begin{equation}
\begin{split}
\sum_{E_n, E_m} &\rho\bigg(\frac{E-E_n}{\hbar}\bigg) 
f_{T}\bigg(\frac{E_m-E_n-\hbar\varepsilon}{\hbar}\bigg)
\abs{\la \psi_n, \Op[a]\psi_m\ra}^2\\
&=\frac{\hat{\rho}(0)}{(2\pi)^d\hbar^{d-1}}\, \int C_E[\sigma(a)](t)\widehat{f_T}(t)\ue^{\ui\varepsilon t}\,\, \ud t
+O\bigg(\hbar^{2-d}  \abs{\rho}_5\abs{a}_{k} \abs{\hat{f}}_0 \frac{1}{\Gamma T}\ue^{\Gamma T}\bigg)\,\, .
\end{split}
\end{equation}
And with the choice \eqref{eq:choice-T} and the rate of weak mixing \eqref{eq:rate-weak-mixing}  
the second relation in 
Theorem \ref{thm:off-diag} follows. 
\end{proof}


\section{Quantum maps}\label{sec:maps}  

In this final section we  study the application of the ideas from the previous sections 
to  the quantisation of some maps on the torus. We will study two classes of maps, 
we begin with perturbed cat maps which are Anosov, and for which we derive the same results 
as for the flows. The second class of examples is given by maps which are ergodic but not hyperbolic. 
This means that we have better control on the remainder term in the Egorov theorem, and in turn 
our method gives for these maps sometimes optimal bounds on the rate of quantum ergodicity.

Let us first quickly review the setup for quantised maps on the 2-torus $T^2=\R^2/\Z^2$, 
see \cite{deB00,DegGra03,MarOKe05} for some 
recent and more complete treatments.  

Instead of one fixed Hilbert space we have now a sequence of Hilbert spaces of dimension $1/\hbar$. 
For each $N\in \N$ the $N$-dimensional Hilbert-space $H_N$ will be identified  with $L^2(\Z_N)$ with the 
inner product 
\begin{equation}
\la \psi,\phi\ra =\frac{1}{N}\sum_{q=1}^N \psi^*(q)\phi(q)\,\, .
\end{equation}
The semiclassical parameter $\hbar$ is identified with $1/N$, so the semiclassical limit is $N\to\infty$.

Operators can be defined again by a Weyl quantisation prescription. For $n=(n_1,n_2)\in\Z^2$ 
define the translation operators on $H_N$ as 
\begin{equation}
T_N(n)\psi(q)=\ue_N(n_1n_2/2) e_N(n_2(q+n_1))\psi(q+n_1)
\end{equation}
where $e_N(x):=\exp(\frac{2\pi \ui}{N} x)$. They satisfy 
\begin{equation}\label{eq:T-prod}
T_N(m)T_N(n)=\ue_N(\omega(m,n)/2) T_N(m+n)\,\, ,
\end{equation}
where $\omega(m,n)=m_1n_2-m_2n_1$, and 
\begin{equation}\label{eq:T-trace}
\Tr T_N(n)=\begin{cases} N & \text{if}\,\,  n=0 \mod N\\ 0 & \text{otherwise}\end{cases}\,\, .
\end{equation}
Now for $a\in C^{\infty}(T^2)$ one defines the Weyl quantisation as 
\begin{equation}\label{eq:torus-weyl}
\Op_N[a]:=\sum_{n\in \Z^2} \hat{a}(n)T_N(n)
\end{equation}
with the Fourier coefficients $\hat{a}(n)=\int_{T^2} a(x)\ue(nx)\,\, \ud x$, where $\ue(x)=\exp(2\pi \ui x)$.

The analogue of Proposition \ref{prop:local-Weyl} is very simple and we state 
it immediately in a form containing products of operators, which we will need later on. 

\begin{lem}\label{lem:map-trace}
Let $a,b\in C^{\infty}(T^2)$, then for all $L\geq 3$ we have 
\begin{equation}
\frac{1}{N}\Tr \Op_N[a]\Op_N[b] =\int_{T^2}a(x)b(x)\,\, \ud x +O\bigg(\frac{\abs{a}_{L}\abs{b}_{3+L}}{N^L}\bigg)\,\, .
\end{equation}
\end{lem}

\begin{proof}
Using the definition \eqref{eq:torus-weyl} and equations \eqref{eq:T-prod}, \eqref{eq:T-trace} we obtain 
\begin{equation}
\begin{split}
\frac{1}{N}\Tr \Op_N[a]\Op_N[b]
&=\sum_{n\in\Z^2}\sum_{m\in\Z}\hat{a}(n)\hat{b}(-n+mN)\ue(\omega(m,n)/2)\\
&=\sum_{n\in\Z^2}\hat{a}(n)\hat{b}(-n)+
\sum_{n\in\Z^2}\sum_{m\in\Z^2\backslash\{0\}}\hat{a}(n)\hat{b}(-n+mN)\ue(\omega(m,n)/2)\,\, .
\end{split}
\end{equation}
Now we have for the first term 
\begin{equation}
\int_{T^2}a(x)b(x)\,\, \ud x=\sum_{n\in\Z^2}\hat{a}(n)\hat{b}(-n)
\end{equation}
and by partial integration for $m\neq 0$
\begin{equation}
\abs{\hat{a}(n)}\leq C \abs{a}_k (1+\abs{n})^{-k}\,\, , \quad 
\abs{\hat{b}(-n+mN)}\leq C \abs{b}_L (1+\abs{n})^L(N\abs{m})^{-L}\,\, ,
\end{equation}
so with $k=L+3$ and $L\geq 3$ the remainder term converges and the result follows.
\end{proof}

The quantisation of a classical volume preserving map $\Phi :T^2\to T^2$ is now defined to be a sequence 
of unitary operators $\{U_N\}_{N\in\N}$ such that for all $a\in C^{\infty}(T^2)$ an Egorov Theorem holds, 
\begin{equation}
\lim_{N\to\infty}\norm{U_N\Op_N[a]U_N^*-\Op_N[a\circ\Phi]} =0\,\, .
\end{equation}
In case that the classical map is ergodic this property allows to prove a quantum ergodicity 
theorem. And as before, in case that we have more detailed information on how the remainder in 
the Egorov Theorem behaves under iteration of the map we can get a bound on the rate of 
quantum ergodicity. We will discuss this now for two examples.

{\bf Perturbed cat maps}: We begin with a class of Anosov maps studied 
recently by Bouclet and De Bi{\`e}vre in \cite{BouDeB04}. 
Let $A\in Sp(2,\Z)$ be a cat map and $g\in C^{\infty}(T^2)$ a real valued function, and consider 
the Hamiltonian flow $\phi^t:T^2\to T^2$ generated by $g$. One can define then 
\begin{equation}
\Phi_{\varepsilon}:=\phi^{\varepsilon}\circ A : T^2\to T^2
\end{equation}
which for small $\varepsilon$ is a small perturbation of the Anosov map $A$, and hence by 
structural stability will be Anosov, too. The quantisation of $\Phi_{\varepsilon}$ is now defined as
\begin{equation}\label{eq:anosov-map}
U_N:=\ue^{-\ui N \varepsilon\Op_N[g]}M_N(A)
\end{equation}
where $M_N(A)$ is the standard metaplectic quantisation of $A$, see, e.g., \cite{deB00,KurRud00,DegGra03}. 
In \cite{BouDeB04} it is now shown that 
there is a constant $\Gamma >0$ such that for $t\in \Z$ 
\begin{equation}\label{eq:Anosov-map-eg}
\norm{U_N^t\Op_N[a]{U_N^*}^t-\Op_N[a\circ\Phi^t_{\varepsilon}]}\leq C_a\frac{1}{N} \ue^{\Gamma \abs{t}}\,\, .
\end{equation}
In fact the estimates in \cite{BouDeB04} are more precise, and $\Gamma$ is estimated quite explicitly, 
but the estimate \eqref{eq:Anosov-map-eg} is sufficient for our purpose.

Using \eqref{eq:Anosov-map-eg} and the trace estimates in Lemma \ref{lem:map-trace} 
we can apply our strategy from the proof 
of Theorem 1 and obtain 

\begin{thm}\label{thm:anosov-map}
Let $U_N$ be the sequence of quantum maps \eqref{eq:anosov-map} and $\psi_j^N$, $j=1,\ldots ,N$ an 
orthonormal basis of eigenfunctions of  $U_N$ for every $N\in\N$. Then for every $a\in C^{\infty}(T^2)$ there is a constant 
$C_a$ such that 
\begin{equation}
\frac{1}{N}\sum_{j=1}^N \abs{\la \psi_j^N, \Op[a] \psi_j^{N}\ra -\overline{a}}^2\leq C_a \frac{1}{\ln N} 
\,\, ,
\end{equation}
where $\overline{a}=\int_{T^2}a \, \ud x$. 
\end{thm}

The same result has been recently proved for the baker's map too, see \cite{DegNonWin04}. 
For cat maps  much stronger results are known, due to their arithmetic nature, 
see \cite{KurRud00,KurRud05}. 

\begin{proof}
We will assume that $\overline{a}=0$. Let $\psi_j^N$, $\ue(\theta_j^N)$, $j=1,\ldots ,N$, be the 
eigenfunctions and eigenvalues of $U_N$, then we have 
\begin{equation}
\Tr \Op_N[a]U_N^{-t} \Op_N[a] U_N^{t}=\sum_{i,j=1}^N \abs{\la \psi_j^N,\Op[a]\psi_i^N\ra}^2 
\ue\big(t(\theta_j^N-\theta_i^N)\big)\,\, . 
\end{equation}
Now choose $f\in \cS(\R)$ such that $\supp \hat{f}\in [-1,1]$, $f\geq 0$ and $f(0)=1$, we have by 
the Poisson summation formula 
\begin{equation}
\sum_{t\in\Z} \frac{1}{T}\hat{f}\bigg(\frac{t}{T}\bigg)\ue\big(t(\theta_j^N-\theta_i^N)\big)
=\sum_{n\in\Z}f\big(T(\theta_j^N-\theta_i^N-n)\big)
\end{equation}
for any $T>0$. By the positivity of $f$ and since $f(0)=1$ we find then 
\begin{equation}
\sum_{i,j=1}^N \abs{\la \psi_j^N,\Op[a]\psi_i^N\ra}^2 
\sum_{t\in\Z} \frac{1}{T}\hat{f}\bigg(\frac{t}{T}\bigg)\ue\big(t(\theta_j^N-\theta_i^N)\big)
\geq \sum_{j=1}^N \abs{\la \psi_j^N,\Op[a]\psi_j^N\ra}^2 \,\, ,
\end{equation}
and so we have the estimate 
\begin{equation}\label{eq:basic-est-maps}
\frac{1}{N}\sum_{j=1}^N \abs{\la \psi_j^N,\Op[a]\psi_j^N\ra}^2
\leq \sum_{t\in\Z} \frac{1}{T}\hat{f}\bigg(\frac{t}{T}\bigg)\frac{1}{N}\Tr \Op_N[a]U_N^{-t} \Op_N[a] U_N^{t}\,\, .
\end{equation}
The Egorov estimate \eqref{eq:Anosov-map-eg} and Lemma \ref{lem:map-trace} give 
\begin{equation}
\begin{split}
\frac{1}{N}\Tr \Op_N[a]U_N^{-t} \Op_N[a] U_N^{t}
&=\frac{1}{N}\Tr \Op_N[a] \Op_N[a\circ \Phi^t_{\varepsilon}] +O\bigg(\frac{\ue^{\Gamma \abs{t}}}{N}\bigg)\\
&=\int_{T^2}a \,a\circ \Phi^t_{\varepsilon}\,\, \ud x 
+O\bigg(\frac{\ue^{3\Gamma' \abs{t}}}{N^3}\bigg)+O\bigg(\frac{\ue^{\Gamma \abs{t}}}{N}\bigg)\,\, ,
\end{split}
\end{equation}
where we have used in addition that there is a constant $\Gamma'>0$ such that 
$\abs{a\circ \Phi^t_{\varepsilon}}_3\leq C \ue^{3\Gamma' \abs{t}}$. Now we can proceed as before 
in the proof of Theorem \ref{thm:rate} with  the choice $T\sim \ln N$ and we use that the 
map $ \Phi^t_{\varepsilon}$ is mixing with an exponential rate, since it is Anosov.   
\end{proof}

An analogue of Theorem \ref{thm:off-diag} could be derived easily with the same methods.

{\bf Parabolic maps}: Our second example will be the parabolic map studied by Marklof and Rudnick 
in \cite{MarRud00}. Let $\alpha\in\R$, then the map $\Psi_{\alpha}: T^2\to T^2$ is defined by 
\begin{equation}\label{eq:par-map}
\Psi_{\alpha}:\begin{pmatrix} p \\ q\end{pmatrix} \mapsto \begin{pmatrix} p+\alpha \\ q+2p\end{pmatrix}\mod 1\,\, .
\end{equation}
If $\alpha$ is irrational this map is uniquely ergodic but not mixing and not hyperbolic. 
This map is quantised in \cite{MarRud00} and it 
is shown that its quantisation $U_N$ satisfies the Egorov estimate 
\begin{equation}\label{eq:par-egorov}
\norm{U_N^{-t}\Op_N[a]U_N^t-Op_N[a\circ \Psi_{\alpha}^t]}\leq C_a\frac{\abs{t}}{N}
\end{equation}
for $t\in\Z$. 

In order to study the rate of quantum ergodicity, we need an estimate on the rate of classical ergodicity.

\begin{lem}\label{lem:rate-of-ergodicity}
Let $a\in C^{\infty}(T^2)$ and $C[a](t)$ be the autocorrelation function 
of the map \eqref{eq:par-map} and assume that $\alpha$ satisfies a Diophantine condition, i.e., there are  
$C,\gamma >0$ such that 
 $\abs{k\alpha -l}\geq C/\abs{k}^{\gamma}$ for all $k,l\in\Z\backslash\{0\}$. 
Then we have for $f\in \cS(\R)$
\begin{equation}\label{eq:erg-weak}
\sum_{t\in\Z}\frac{1}{T}\hat{f}\bigg(\frac{t}{T}\bigg)C[a](t)=O\bigg(\frac{1}{T}\bigg)\,\, ,
\end{equation}
where $\hat{f}$ denotes the Fourier-transform of $f$.  Furthermore, if $a$ depends only on $p$ then 
\begin{equation}\label{eq:erg-strong}
\sum_{t\in\Z}\frac{1}{T}\hat{f}\bigg(\frac{t}{T}\bigg)C[a](t)=O_M\bigg(\frac{1}{T^M}\bigg)\,\, , \quad \text{for all}\,\, M\in\N\,\, .
\end{equation}
\end{lem}

\begin{proof}
We have 
\begin{equation}
\Psi_{\alpha}^t:\begin{pmatrix}p \\ q\end{pmatrix}\mapsto\begin{pmatrix} p+t\alpha \\ q+2tp +\alpha t(t-1)\end{pmatrix}\,\, , 
\end{equation}
and with $a(x)=\sum_{n\in\Z^2} \hat{a}(n)\ue(nx)$ we get 
\begin{equation}
C[a](t)=\sum_{n,m\in\Z^2\backslash\{0\}} \hat{a}(n)\hat{a}(m)\int_{T^2}\ue(nx)\ue\big(m\Psi_{\alpha}^t(x)\big)\,\, \ud x\,\, .
\end{equation}
Then we find 
\begin{equation}
\int_{T^2}\ue(nx)\ue\big(m\Psi_{\alpha}^t(x)\big)\,\, \ud x
=\delta(-n_1,m_1+2tm_2)\delta(-n_2,m_2) \ue(m_1 \alpha t +m_2 \alpha t(t-1))\,\, ,
\end{equation}
where $\delta(m,n)$ denotes the Kronecker delta, and therefore 
\begin{equation}
C[a](t)=\sum_{(m_1,m_2)\in\Z^2\backslash\{0\}} \hat{a}(-m_1-2tm_2,-m_2)\hat{a}(m_1,m_2)\ue(m_1 \alpha t +m_2 \alpha t(t-1))\,\, .
\end{equation}
Now we split $C[a](t)$ into two parts, $C[a](t)=C^0[a](t)+C^1[a](t)$, such that $C^0[a](t)$ contains only the 
terms with $m_2=0$
\begin{equation}
C^0[a](t)=\sum_{m\in\Z\backslash\{0\}} \hat{a}(-m,0)\hat{a}(m,0)\ue(m \alpha t)\,\, .
\end{equation}
The second term satisfies 
\begin{equation}
\abs{C^1[a](t)}\leq C_K (1+\abs{t})^{-K}
\end{equation}
for all $K\in\N$ since the Fourier-coefficients $\hat{a}(n)$ are quickly decreasing and  
therefore 
\begin{equation}
\sum_{t\in\Z}\frac{1}{T}\hat{f}\bigg(\frac{t}{T}\bigg)C^1[a](t)=O\bigg(\frac{1}{T}\bigg)\,\, .
\end{equation}
For the first term we find 
\begin{equation}
\sum_{t\in\Z}\frac{1}{T}\hat{f}\bigg(\frac{t}{T}\bigg)C^0[a](t)
=\sum_{m\in\Z\backslash\{0\}} \abs{\hat{a}(m,0)}^2
\sum_{t\in\Z}\frac{1}{T}\hat{f}\bigg(\frac{t}{T}\bigg)\ue(m \alpha t)
\end{equation}
and by the Poisson summation formula we obtain 
\begin{equation}
\sum_{t\in\Z}\frac{1}{T}\hat{f}\bigg(\frac{t}{T}\bigg)\ue(m \alpha t)=
\sum_{n\in\Z}f(T(m \alpha-n))=O_M(\abs{m}^{\gamma M}T^{-M})
\end{equation}
since $f\in\cS(\R)$ and by the Diophantine condition on $\alpha$. And since 
the Fourier-coefficients $\hat{a}(n)$ are quickly decreasing we find 
\begin{equation}
\sum_{t\in\Z}\frac{1}{T}\hat{f}\bigg(\frac{t}{T}\bigg)C^0[a](t)
=O_M(T^{-M})
\end{equation}
Combining the two estimates for $C^0[a](t)$ and $C^1[a](t)$ gives the lemma. 
\end{proof}

Combining the Egorov estimate and this lemma we then obtain

\begin{thm}\label{thm:parabolic}
Let $U_N$ be the quantisation of the map \eqref{eq:par-map} due to 
\cite{MarRud00} with a Diophantine $\alpha$, and $\psi_j^N$, $j=1,\ldots N$, a 
orthonormal basis of eigenfunctions.  Then we have 
\begin{equation}\label{eq:slow-qe}
\frac{1}{N}\sum_{j=1}^N\abs{\la \psi_j^N,\Op[a]\psi_j^N\ra-\bar{a}}^2\leq C_a\frac{1}{N^{1/2}}\,\, ,
\end{equation}
and if $a$ depends on $p$ only then we have the stronger estimate 
\begin{equation}\label{eq:fast-qe}
\frac{1}{N}\sum_{j=1}^N\abs{\la \psi_j^N,\Op[a]\psi_j^N\ra-\bar{a}}^2\leq C_{a,\varepsilon}\frac{1}{N^{1-\varepsilon}}\,\, ,
\end{equation}
for every $\varepsilon>0$. 
\end{thm}

\begin{proof}
Using the estimate \eqref{eq:basic-est-maps}   from the proof of Theorem \ref{thm:anosov-map} we have to 
estimate 
\begin{equation}
\sum_{t\in\Z} \frac{1}{T}\hat{f}\bigg(\frac{t}{T}\bigg)\frac{1}{N}\Tr \Op_N[a]U_N^{-t} \Op_N[a] U_N^{t}\,\, .
\end{equation}
Now $\abs{a\circ \Psi_{\alpha}^t}_k\leq C_k \abs{t}^k$, so with Lemma \ref{lem:map-trace} and the 
Egorov estimate \eqref{eq:par-egorov} we get 
\begin{equation}
\begin{split}
\frac{1}{N}\Tr \Op_N[a]U_N^{-t} \Op_N[a] U_N^{t}
&=\frac{1}{N}\Tr \Op_N[a] \Op_N[a\circ \Psi_{\alpha}^t]+O\bigg( \frac{\abs{t}}{N}\bigg) \\
&=C[a](t)+O\bigg( \frac{\abs{t}^3}{N^3}\bigg) +O\bigg( \frac{\abs{t}}{N}\bigg) \,\, .
\end{split}
\end{equation}
If we use then \eqref{eq:erg-weak} we obtain 
\begin{equation}
\sum_{t\in\Z} \frac{1}{T}\hat{f}\bigg(\frac{t}{T}\bigg)\frac{1}{N}\Tr \Op_N[a]U_N^{-t} \Op_N[a] U_N^{t}
=O\bigg(\frac{1}{T}\bigg)+O\bigg( \frac{T^3}{N^3}\bigg) +O\bigg( \frac{T}{N}\bigg)\,\, ,
\end{equation}
and so the choice $T=N^{1/2}$ gives \eqref{eq:slow-qe}. If we have instead the faster decay 
\eqref{eq:erg-strong} we get 
\begin{equation}
\sum_{t\in\Z} \frac{1}{T}\hat{f}\bigg(\frac{t}{T}\bigg)\frac{1}{N}\Tr \Op_N[a]U_N^{-t} \Op_N[a] U_N^{t}
=O_M\bigg(\frac{1}{T^M}\bigg)+O\bigg( \frac{T^3}{N^3}\bigg) +O\bigg( \frac{T}{N}\bigg)\,\, ,
\end{equation}
for every $M\in\N$ and so by choosing $T=N^{\varepsilon'}$, with $\varepsilon'$ small enough, 
and $M$ large enough we obtain \eqref{eq:fast-qe}.
\end{proof}

The results in \cite{MarRud00} show that the estimate \eqref{eq:slow-qe} is optimal, so in this case we 
obtain a sharp estimate. The analysis in \cite{MarRud00} is much more detailed and they have sharp estimates 
for the rate of quantum ergodicity for individual eigenfunctions. But Theorem \ref{thm:parabolic} 
might still be of some interest because the proof is of a more dynamical nature, and therefore may be easier 
to extend to more general cases.

One further class of systems where one could apply the same methods is given by perturbed Kronecker maps, 
which were  recently studied by Rosenzweig, \cite{Ros05}. Here the proof would be very similar to the one 
of \eqref{eq:fast-qe}, and we would get the same rate $O_{\varepsilon}(1/N^{1-\varepsilon})$. But in 
\cite{Ros05} an stronger bound on individual eigenfunctions is given, so our method does not give 
an optimal result.

The results of Theorem \ref{thm:off-diag} do not hold for these maps since they are not weakly mixing. 
In particular, using the same methods as in the proof of Lemma \ref{lem:rate-of-ergodicity}, one finds for 
$\Psi_{\alpha}^t$ that for 
$\varepsilon =k\alpha$, $k\in \Z\backslash\{0\}$, 
\begin{equation}
\sum_{t\in\Z}\frac{1}{T}\hat{f}\bigg(\frac{t}{T}\bigg)C[a](t)\ue(\varepsilon t)=\abs{\hat{a}(k,0)}^2+O\bigg(\frac{1}{T}\bigg)\,\, .
\end{equation}
From this result together with the techniques used in the proof of Theorem \ref{thm:parabolic} one can derive 
\begin{equation}
\lim_{N\to\infty} \frac{1}{N} \sum_{\abs{\theta_i-\theta_j- \varepsilon/N}\leq 1/N^{1/2}}^N \abs{\la \psi_i^{N}, \Op_N[a]\psi_j^N\ra}^2 
=\abs{\hat{a}(k,0)}^2\,\, , 
\end{equation}
where $\psi_i$, $\ue(\theta_i)$, $i=1, \ldots ,N$  are the eigenvectors and eigenvalues of 
$U_N$, and  $\varepsilon =k\alpha$. So weak mixing is a necessary condition for the validity of 
\eqref{eq:off-diag-weak-mix} in Theorem \ref{thm:off-diag}.

{\footnotesize

\newcommand{\etalchar}[1]{$^{#1}$}
\def\cprime{$'$}
\providecommand{\bysame}{\leavevmode\hbox to3em{\hrulefill}\thinspace}
\providecommand{\MR}{\relax\ifhmode\unskip\space\fi MR }
\providecommand{\MRhref}[2]{%
  \href{http://www.ams.org/mathscinet-getitem?mr=#1}{#2}
}
\providecommand{\href}[2]{#2}

}

\end{document}